# Femtosecond laser carbonization of polystyrene

Andrey Kudryashov[1], Sergey Gusev[2], Anastasiya Orlova[2], Andrey Afanasiev[1], Maria Sveshnikova[1], Alexander Pikulin[1], and Nikita Bityurin[1,*]

[1]Institute of Applied Physics of the Russian Academy of Sciences, 46 Ul'yanov Street, 603950 Nizhny Novgorod, Russia

[2]Institute of Physics of Microstructures of the Russian Academy of Sciences, 603950, Nizhny Novgorod GSP-105, Russia

*bit@appl.sci-nnov.ru

## Abstract

Multi-pulse femtosecond laser irradiation of a monolayer of polystyrene microspheres deposited on a polystyrene substrate leads to the formation of carbon nanomaterial exhibiting broadband excitation-dependent luminescence both within the microspheres and in the substrate. Initial polystyrene substrate and microspheres are transparent at the laser wavelength (800 nm). Peak intensity of the laser irradiation focusing by the microspheres reaches $10^{13}$ W/cm$^2$, resulting in multiphoton absorption followed by ionization and further carbonization processes. Raman spectroscopy and transmission electron microscopy analysis show that carbonization products contain carbon crystalline nanoobjects.

## Introduction

Composite materials consisting of polymer matrix with embedded nanoparticles of different nature are promising materials for numerous applications due to their unique optical, electrical, chemical and mechanical properties [1-4]. Nanoparticles (NPs) can be synthesized directly within the polymer matrix as a result of photothermal/ photochemical decomposition of the precursor molecules [5-7]. This allows direct laser writing of nanoparticle based patterns, opening up an approach to printing optoelectronic devices using a focused laser beam [8] or a mask of colloidal microspheres [9].

Among the diversity of materials for nanoparticle manufacturing, carbon has recently drawn increasing interest. Carbon is a widespread material, it has low toxicity, and carbon nanoobjects exhibit extraordinary optical and electrical properties, which makes them extremely useful material in such fields as optoelectronics, bioimaging, energy storage, etc [10-13]. Different types of carbon dots demonstrate bright tunable luminescence [11, 14, 15]. In [16], the authors report carbon dots with a broad emission spectrum in the visible region. These types of emitting sources find applications for white LEDs [17].

In contrast to photoinduced nanocomposites with semiconductor or metal nanoparticles, where a soluble photosensitive precursor is required [5-7], carbon nanoobjects can be obtained directly from the polymer matrix itself by carbonization. Laser carbonization of solid polymers is well studied in numerous publications [18-21]. Laser induced graphene with an excellent electrical conductivity can be obtained by irradiation of a polyimide film [19]. In [20], the authors demonstrate the simultaneous synthesis and patterning of luminescent carbon dots inside a polydimethylsiloxane matrix via laser direct writing.

A promising matrix for laser-induced nanocomposites, including laser carbonization, is polystyrene (PS) [22], which has good thermomechanical properties, optical transparency and low cost. Recent publication reports gold nanoparticle embedded PS carbonization after irradiation with nanosecond laser pulses at the wavelength of the plasmon resonance [23]. Here, luminescent nanostructures of disordered carbon appear directly near the gold NPs, where the local heating reaches 2000 K. Exposure of pure polystyrene microspheres with femtosecond UV pulses of the third harmonic of a Ti:Sa laser (266 nm) results in the appearance of white luminescence [24]. The authors attribute this phenomenon to the formation of carbonyl groups by photochemical oxidation rather than carbonization. In [25], it was shown that irradiation of toluene, which has similar chemical structure as PS, with the high power laser pulses of the fundamental frequency of a Ti:Sa laser leads to the formation of luminescent carbon NPs. The authors suggest that carbonization is driven by photoionization and the formation of an electron plasma in liquid.

In this paper, we study femtosecond laser (fundamental frequency (FF) of a Ti:Sa laser) polystyrene carbonization. Irradiation was carried out through a layer of polystyrene microspheres deposited on a polystyrene substrate. The mask of microspheres acts like an array of lenses and was previously used for patterning of luminescent jets consisting of semiconductor nanoparticles [9]. We discovered that carbonization occurs both in the PS substrate and in the microspheres. Carbonization products exhibit photoluminescence with a broad emission spectrum. The intensity of incident radiation focused by microspheres reaches $10^{13}$ W/cm$^2$. In contrast to the disorder carbon in [23] formed by local heating, carbon nanoobjects obtained here are crystalline.

## Materials and methods

A drop of a 5-μm polystyrene microsphere aqueous suspension (Lenkhrom, Russia) was deposited on a polymer substrate. The drop then dried forming a close-packed monolayer of microspheres. Two different polymer substrates were used, PS and polymethylmethacrylate (PMMA).

Irradiation of the samples was performed using the FF of a Ti:Sa laser (800 nm) with a pulse duration $\tau$ = 30 fs ($I = I_0 e^{-t^2/\tau^2}$) and a pulse energy of 0.6 mJ (Spitfire Pro laser system with Tsunami master oscillator). Both the microspheres and the substrate are transparent at a wavelength of 800 nm. The intensity of the laser beam, at which the sample blackens and the microspheres are not removed from the substate, was adjusted using a converging lens (fig. 1(a)). The corresponding fluence was 8.5 mJ/cm$^2$ (peak intensity 1.6·10$^{11}$ W/cm$^2$) and the pulse repetition rate was 100 Hz. The samples were irradiated for 5 minutes.

A setup for photoluminescent measurements is shown in fig. 1(b). Luminescence was excited by a 405 nm continuous-wave laser (Lasever Inc., Ningbo, China). To study the dependence of luminescence on the excitation wavelength, a pulsed nanosecond Ti:Sa laser with a tunable wavelength (LT-2211N, Lotis TII, Minsk, Belarus) operated at the second harmonic (350-450 nm) was applied. A converging lens focuses the luminescent radiation into the input of the optical fiber of the Ocean Optics QE65Pro spectrometer (Ocean Optics, Dunedin, FL, USA). An ET-425LP sharp-edged low-pass spectral filter (Chroma Technology Corp, Bellows Falls, VA, USA) rejects scattered excitation light.

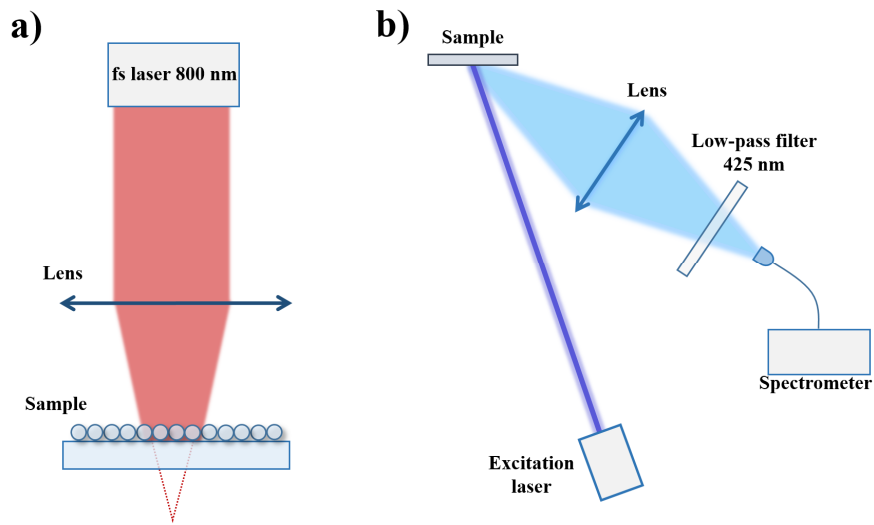

Fig. 1. An experimental setup for laser irradiation (a), and for luminescence measurements (b).

Transmission electron microscopy (TEM) has been used to characterize the nanoparticle size, shape, structure and aggregation. TEM measurements were carried out with the LIBRA 200 MC (Carl Zeiss AG, Germany) operated at a voltage of 200 kV and having a resolution limit of 0.12 nm. The TEM images were acquired and processed using Digital Micrograph software (Gatan, Inc., USA).

Raman spectra were measured using Raman spectrometer Bruker Senterra (Optik GmbH, Germany). An excitation wavelength was 785 nm, a laser power was 1 mW, microscope objective was x10, NA = 0.25.

Large-scale microscopic images were obtained with Nikon Eclipse Ci-S microscope (Japan), equipped with an NVSU233A UV LED (Nichia, Japan) for luminescent image acquisition. An input filter in the microscope transmits radiation in the range of 380-400 nm. An output filter was the same as in the luminescent spectra measurements setup (fig. 1.b).

Numerical calculation of the light intensity near the monolayer of microspheres was performed using the FDTD method with periodic boundary conditions as described in [26].

The scans of the surface of the irradiated samples were provided using semi-contact mode of the Ntegra atomic-force microscope (AFM) (NT-MDT, Russia).

## Results and discussions

Microscopic image of the monolayer of the microspheres on PS substrate is shown in Fig.2.

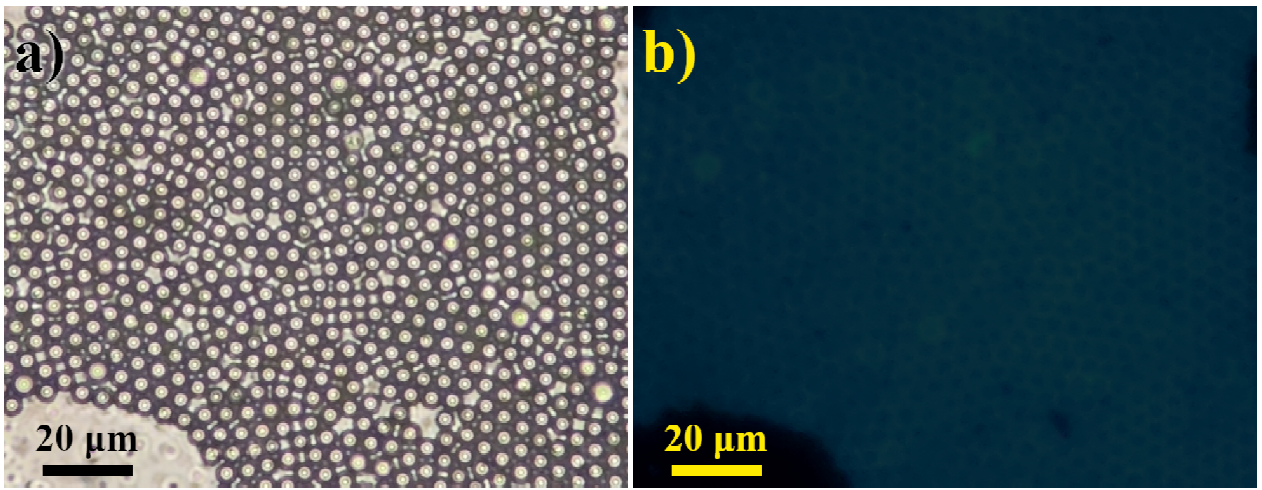

Fig. 2. Microscopic image of the sample (PS microspheres on PS substrate) before femtosecond laser irradiation when illuminated with white light (a) and UV (b).

Non-irradiated PS microspheres exhibit low-intensity, barely noticeable, blue luminescence (Fig. 2(b)).

Femtosecond laser irradiation causes blackening of the sample. Figure 3 shows the microscopic images of the irradiated sample before the microspheres were removed.

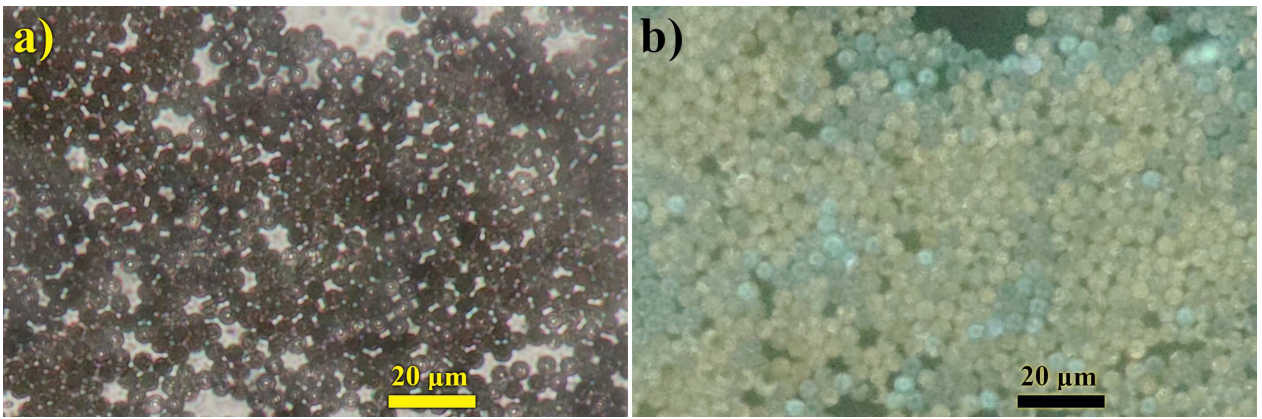

Fig. 3. Microscopic image of the irradiated sample (PS microspheres on PS substrate) before removing of the microspheres when illuminated with white light (a) and UV (b).

Significant difference of the non-irradiated (Fig. 2(a)) and irradiated (Fig. 3(a)) sample is observed, irradiated microspheres are black due to the products of carbonization. At the same time, these products demonstrate broadband, almost white, photoluminescence when illuminated with UV (Fig. 3.b).

Figure 4 demonstrates the microscopic images of the irradiated sample after the microspheres have been removed.

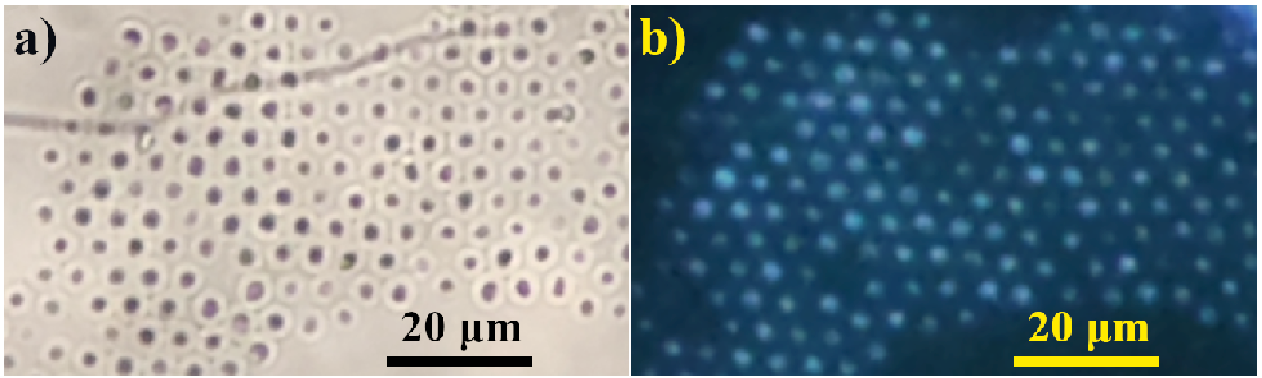

Fig. 4. Microscopic image of the irradiated sample (PS microspheres on PS substrate) after removing of the microspheres when illuminated with white light (a) and UV (b).

Under the removed spheres, there are traces in the PS matrix, which are also contain carbonizations products and exhibit luminescence. Although, the carbonization effect is much greater in the PS spheres themselves, than in PS matrix under the spheres (Fig. 3, Fig. 4) there is still possible laser patterning of luminescent microstructures.

Figure 5 shows AFM images of the irradiated sample (the surface of PS substrate after removing of the microspheres). As the result of irradiation, craters are formed under the microspheres. Those craters contain luminescent products (Fig. 4).

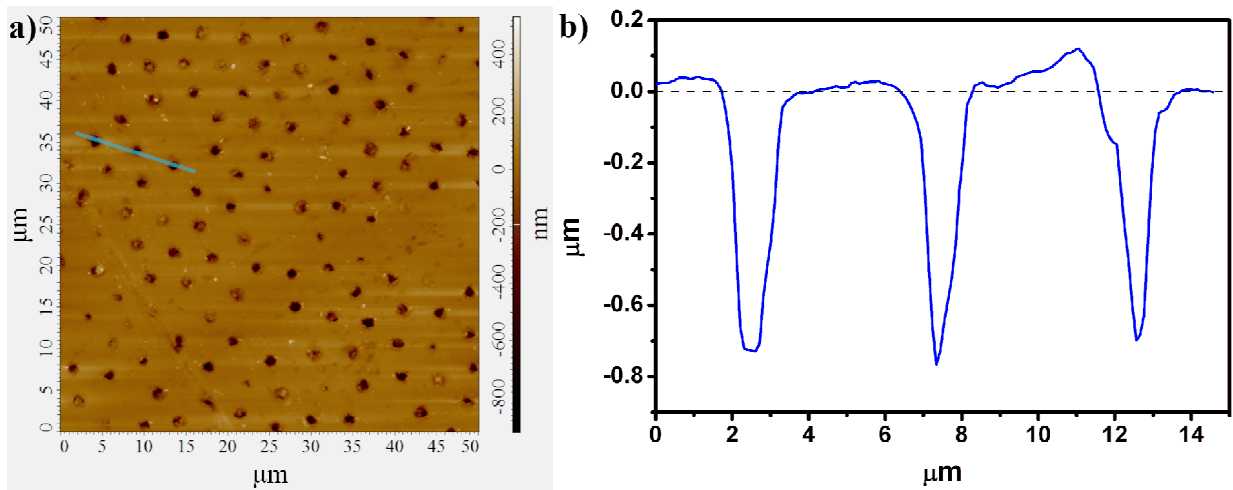

Fig. 5. (a) AFM scan of the surface of the irradiated sample (PS microspheres on PS substrate) after removing of the microspheres. (b) Profile along the blue line in (a).

Figure 6 shows the results of numerical calculation of the distribution of laser intensity in the area of polystyrene microspheres on a polystyrene substrate (refractive index of PS $n = 1.58$ at a wavelength of 800 nm).

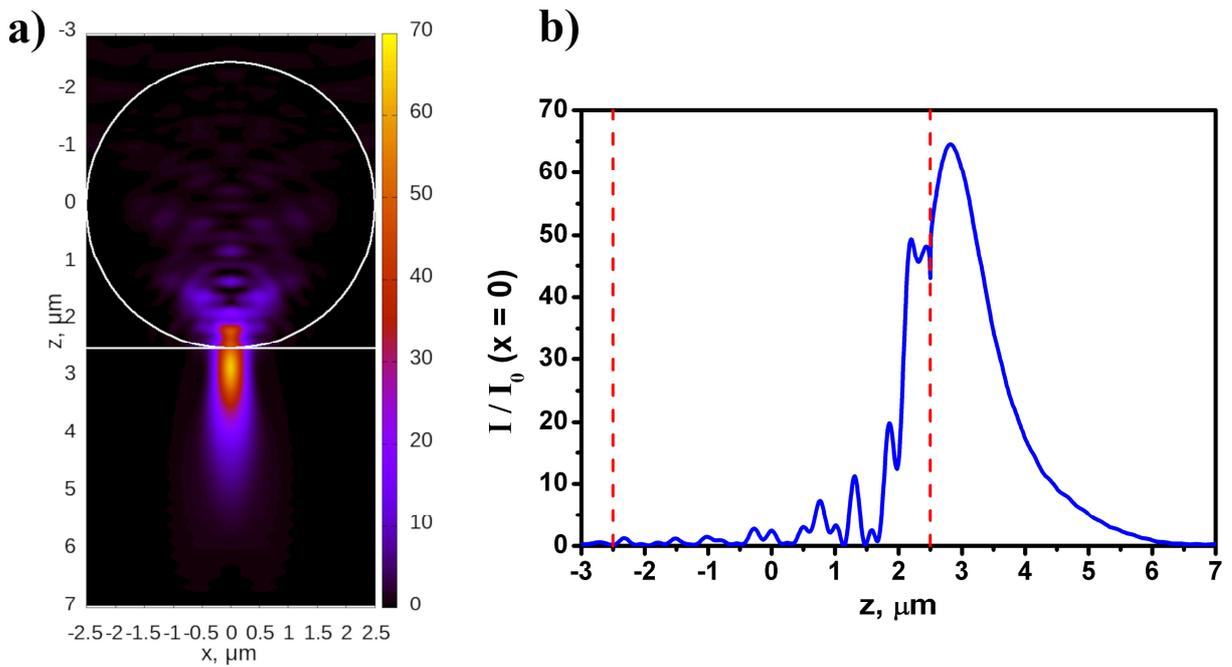

Fig. 6. FDTD calculation of the laser intensity normalized to the input intensity, in the microsphere and in the substrate. The refractive index of both the sphere and the substrate $n = 1.58$. The sphere diameter is 5 μm, the wavelength is 800 nm. (a) Total intensity distribution, (b) intensity distribution along the optical axis.

As it is seen in Fig. 6, the maximum intensity is achieved in the substrate, but the radiation intensity within the sphere at the boundary with the substrate is also quite high. Note that the peak intensity of laser radiation, taking into account the focusing by the sphere, is about $10^{13}$ W/cm$^2$. Thus, carbonization at the beginning of irradiation (irradiation is fundamentally multi-pulse) occurs in the substrate, then carbonization proceeds inside the sphere. Since this process leads to a significant increase in the absorption coefficient at the wavelength of irradiation, this radiation is intercepted inside the sphere and does not reach the substrate. That is, after some time, carbonization occurs only in the spheres and provides a sufficient number of products for analysis. Note that the process occurs simultaneously in a very large number ($2 \cdot 10^6$) of the spheres.

Figure 7 shows the luminescence spectra of the PS microspheres on PS substrate, when excited at 405 nm.

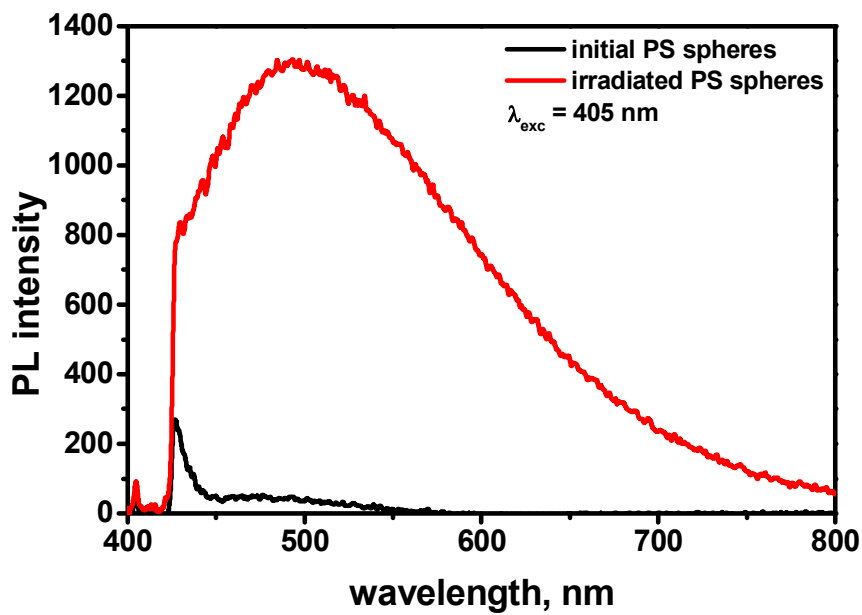

Fig. 7. Luminescence spectra of the sample excited at a wavelength of 405 nm.

Irradiation of the spheres leads to significant increase of luminescence signal. Irradiated sample has a broad luminescence spectrum over the entire visible range. To study the dependence of luminescence of the irradiated sample on the excitation wavelength, a tunable laser was applied. The measurements were carried out in the excitation wavelength range of 370–420 nm (fig. 8).

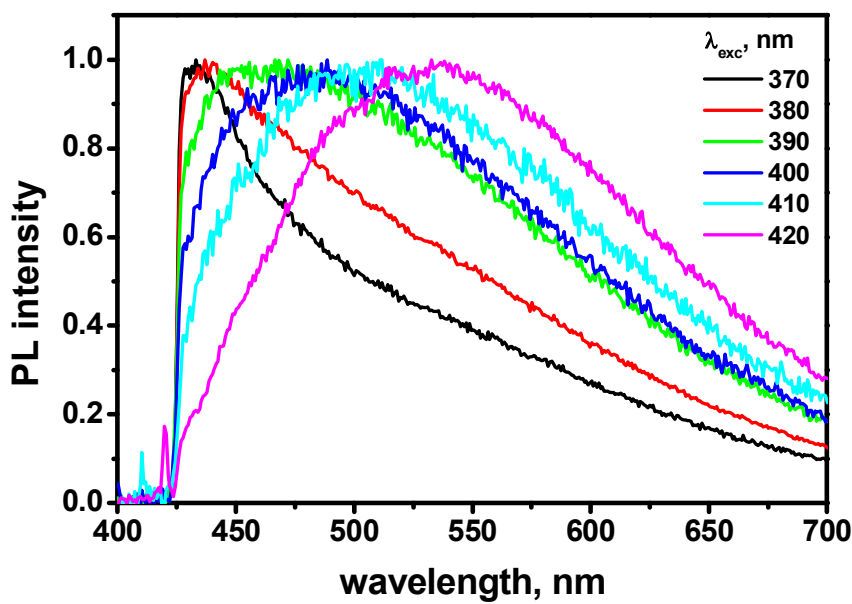

Fig. 8. Luminescence spectra of the irradiated spheres on PS substrate at different excitation wavelengths (370 – 420 nm, 2nd harmonic of tunable Ti:Sa laser).

Figure 8 shows that the luminescence spectrum shifts toward longer wavelengths with increasing excitation wavelength. This behavior is characteristic of different types of luminescent carbon nanostructures [14, 15, 23]. A similar phenomenon was observed in our previous work, where disordered carbon was obtained via gold NP mediated nanosecond laser heating [23].

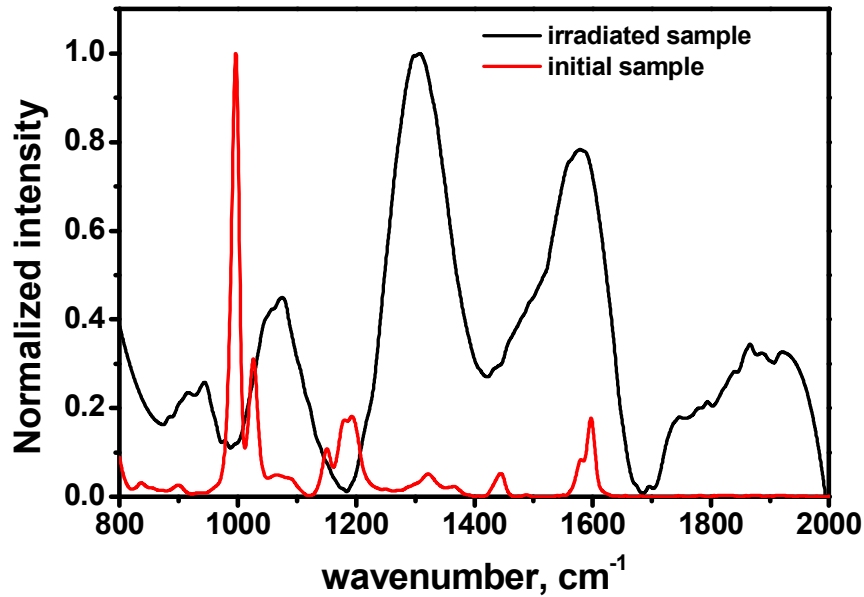

Fig. 9. Raman spectra of the PS spheres on PS substrate before and after femtosecond laser irradiation.

Raman spectra of the sample (PS microspheres on PS substrate) before and after femtosecond laser exposure in Fig. 9 shows that under the effect of a laser irradiation there is a disappearance of the characteristic PS band around 1000 cm$^{-1}$. In the irradiated sample, bands can be clearly identified in the spectrum in the regions G (~1580 cm$^{-1}$) and D (~1300 cm$^{-1}$). G peak corresponds to vibrations of sp2 aromatic rings, D peak can be attributed to microcrystalline graphitic sheets [27, 28].

For TEM study the irradiated polymer film was dissolved in toluene. Because of this procedure, a dark sediment appeared at the bottom of the container with the solution. This sediment was washed several times in fresh toluene. After that, the washed sediment was dispersed in a small amount of acetone using ultrasound. A small drop of the resulting suspension was dripped onto a 50 nm silicon nitride membrane ($Si_3N_4$) and, after complete drying, we examined the resulting sample under an electron microscope.

The nanoparticles size was estimated manually based on the Bright Field (BF) and the Dark Field (DF) TEM images. TEM studies have shown that the particles have a variety of morphologies and their sizes vary widely. The BF image (Fig. 10(a)) clearly shows particles (dark formations of round and irregular shape) with lateral sizes from 50 nm to 1 μm and more. However, in the DF image (Fig. 10(b)) we observe that the coherent scattering regions have much smaller sizes, approximately from 2 to 5 nm. Based on this, it can be assumed that large particles are agglomerations of nanoparticles with dimensions of several nm. This is also likely the reason why we were unable to obtain high-resolution images of the nanoparticles.

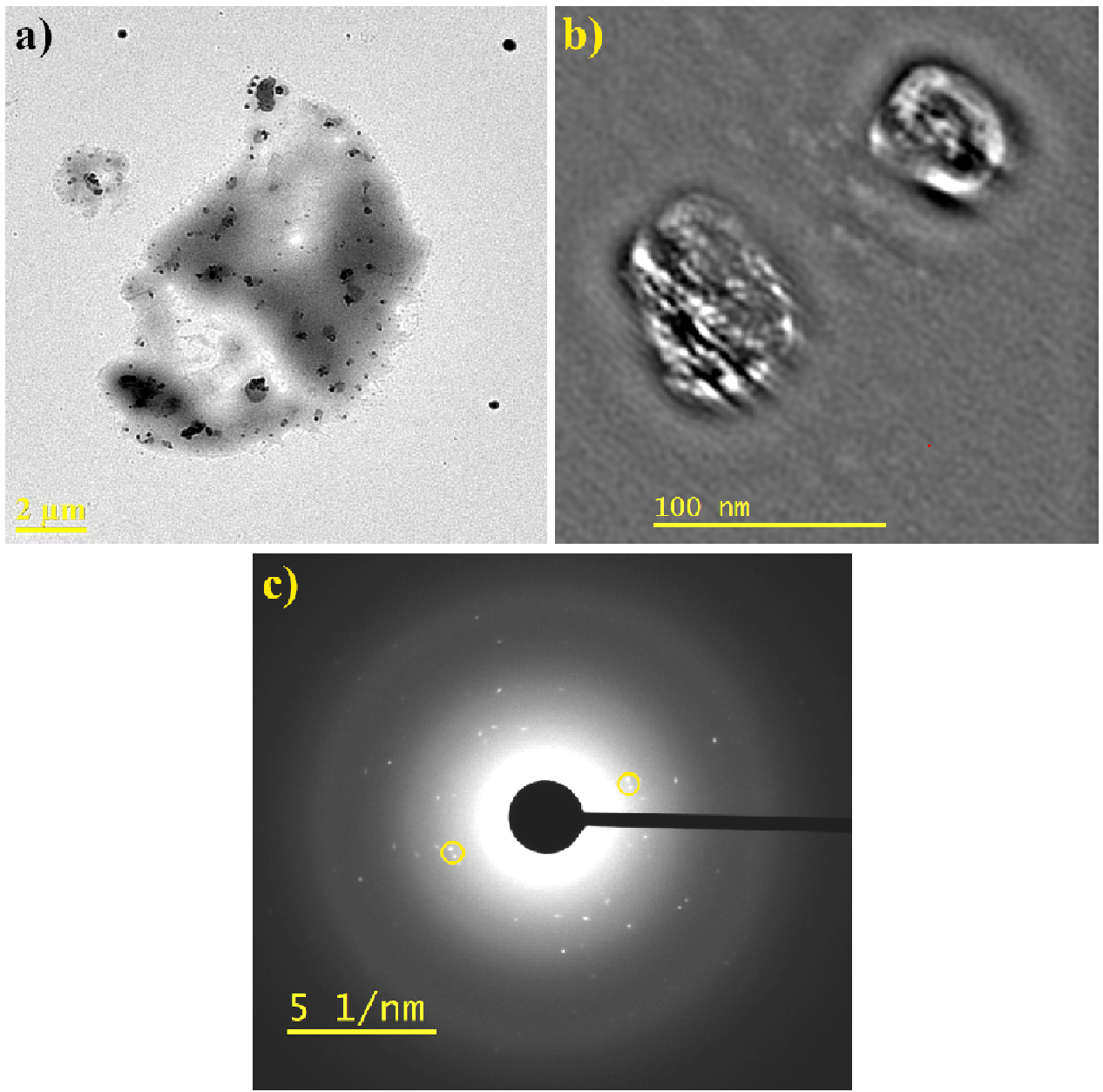

Fig. 10. (a) TEM Bright Field micrograph of the particles on the Si$_3$N$_4$ membrane. The particles are seen as dark spots. (b) TEM Dark Field micrograph of two particles. The carbon nanoparticles are visible as light dots. (c) SAD image of one agglomeration of nanoparticles. Unrecognized patterns are marked with yellow circles.

It is well known that the carbon structure can have a variety of allotropic forms [29]. In the Selected Area Diffraction (SAD) image we see a large number of patterns (Fig. 10(c)). Almost all reflections correspond to two structural phases of carbon Cmmm and R-3m. Unfortunately, it was not possible to identify two strong reflections, which were marked with a yellow circle in Fig. 10(c) and were obtained from a crystal lattice with a parameter of 0.303±0.0005 nm. The carbon structure with such parameters is aabsent in the database available to us [29].

TEM analysis, Raman spectroscopy, characteristic luminescence spectra, and the shift of the luminescence band with a change in the excitation wavelength, suggest that as a result of multi-pulse femtosecond laser exposure, crystalline carbon nanoparticles are formed in polystyrene microspheres and substrate.

An additional experiment (under the same conditions) was carried out with PS microspheres on a PMMA substrate instead of PS. In this case, PS spheres were also carbonized after irradiation and demonstrated the same luminescence properties (Fig. 11).

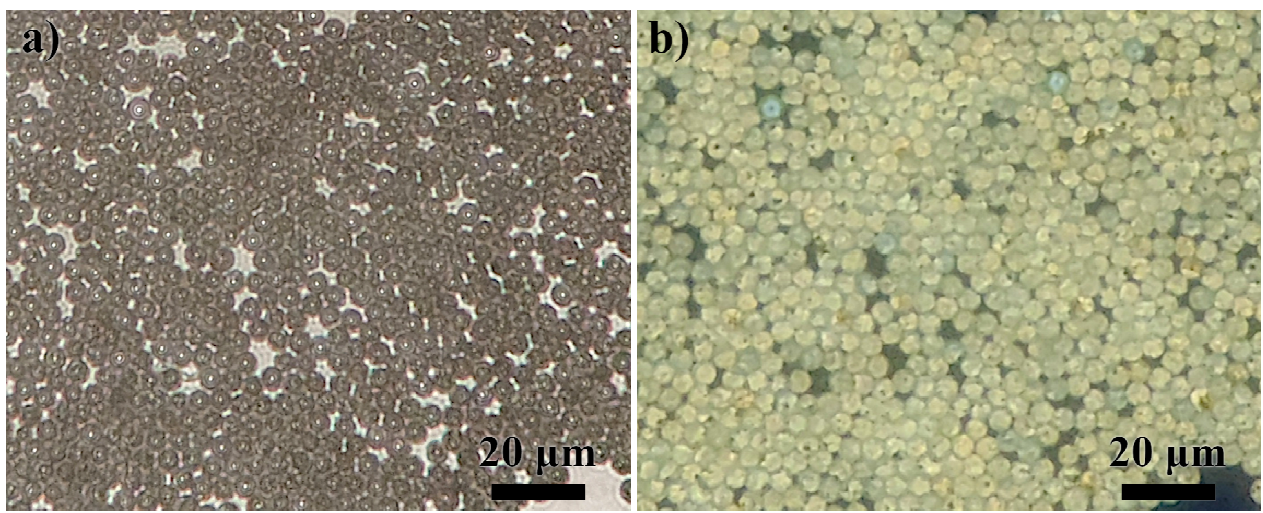

Fig. 11. Microscopic image of the irradiated sample (PS microspheres on PMMA substrate) before removing of the microspheres when illuminated with white light (a) and UV (b).

The photograph in Fig. 12(a) was taken at the border of the irradiated area to show the contrast between irradiated and initial spheres. Carbonized spheres (bottom) exhibit bright yellow-green luminescence and non-irradiated spheres (top) weakly luminescence with blue light.

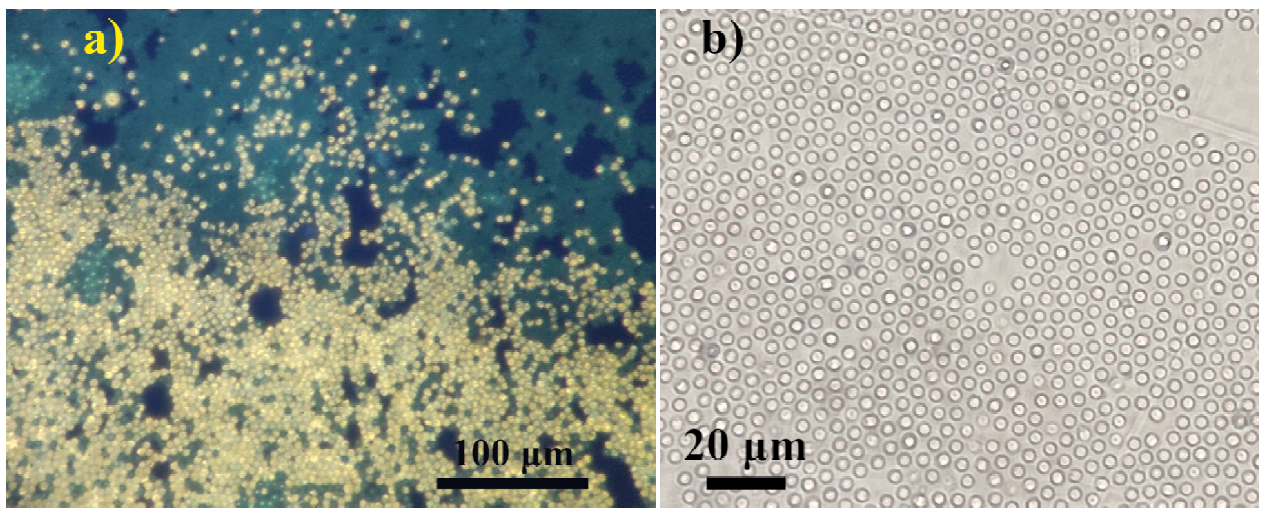

Fig. 12. (a) Luminescent microscopic image of the sample (PS microspheres on PMMA substrate) obtained with UV illumination. (b) Microscopic image of the surface of the irradiated sample (PMMA substrate after microspheres were removed).

A microscopic image of the surface of PMMA after irradiation and removing of the microspheres is shown in Fig. 12(b). There are traces on the substrate under the spheres, but, unlike traces on PS substrate (Fig. 4), they do not luminesce when probed by UV.

AFM image of the surface of PMMA substrate (Fig. 13) show ablation craters, which are narrower then craters on PS substrate (Fig. 5) and in contrast to the craters on PS they do not contain luminescent products of carbonization.

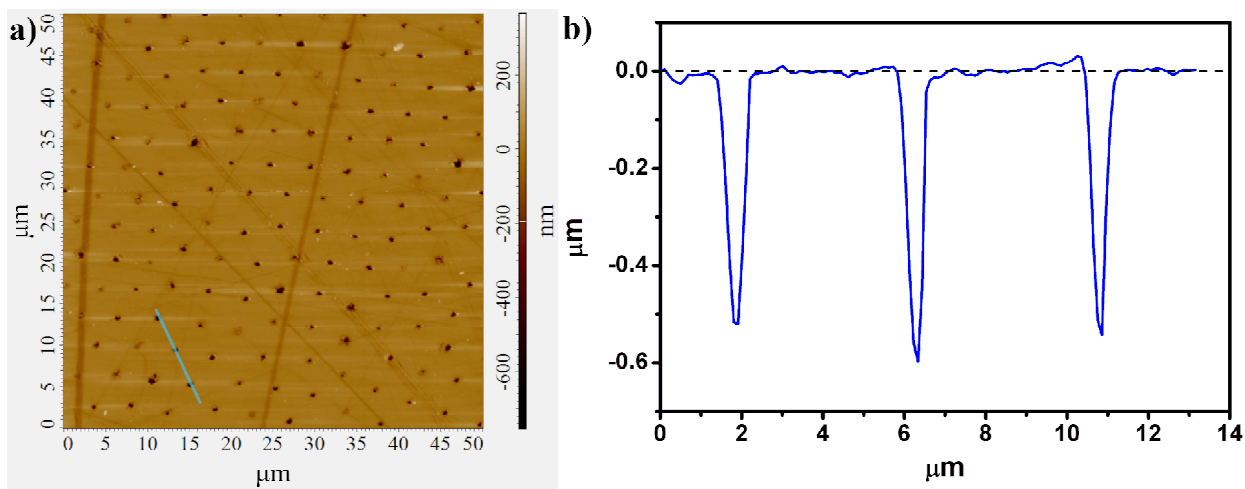

Fig. 13. (a) AFM scan of the surface of the irradiated sample (PS microspheres on PS substrate) after removing of the microspheres. (b) Profile along the blue line in (a).

## Conclusion

The samples of monolayer of PS microspheres on the polymer (PS and PMMA) substrates, where microspheres act like array of lenses, were exposed to femtosecond 800 nm laser pulses. FDTD calculations show that the microspheres focus laser radiation both into the substrate and inside themselves. Peak intensity inside microspheres reaches $10^{13}$ W/cm$^2$, which results in multiphoton absorption followed by ionization with further carbonization.

Laser carbonization cause blackening of the samples and appearance of the broadband (almost white) luminescence. When the substrate is PS, carbonization and luminescent products occur both in the microspheres and in the substrate and laser patterning of luminescent microstructures is possible.

Carbonization products were studied using TEM, Raman, and luminescence spectroscopy. Measurement the luminescence at different excitation wavelengths shows that the luminescence spectrum shifts toward longer wavelengths with increasing excitation wavelength, which is typical for luminescent carbon nanoobjects. Raman spectra contain D and G bands, which confirms the formation of carbon in the irradiated samples. TEM analysis shows the presence of crystalline nanoparticles with sizes of several nanometers consisting of two structural phases of carbon Cmmm and R-3m.

Carbon obtained in present paper is crystalline, as it was in [25], where toluene was irradiated with 800 nm femtosecond laser. This significantly differs from disordered carbon obtained in our previous work on polystyrene carbonization by means of nanosecond laser heating of embedded gold nanoparticles [23]. It allows suggesting that carbonization here is mediated by plasma relevant processes.

*Funding*

This work was supported by the Russian Science Foundation under project No. 22-19-00322. and the Ministry of Science and Higher Education of the Russian Federation, project FFUF-2024-0030

*Acknowledgments*

This work was performed under project 22-19-00322 RSF except FDTD calculations and AFM measurements, which carried out under project FFUF-2024-0030 of the Ministry of Science and Higher Education of the Russian Federation

The Raman spectroscopy studies were performed using the equipment of the Center for Collective Use at the Institute of Chemical Physics Russian Academy of Sciences, Moscow, Russian Federation.

*Disclosures*

The authors declare no conflicts of interest.

*Data availability*

No additional data were generated or analyzed in the presented research.